\documentclass[a4paper,pra,showpacs,twocolumn]{revtex4}
\usepackage{amsfonts}
\usepackage{amssymb}
\usepackage{amsmath}
\usepackage{calc}
\usepackage{graphicx}
\usepackage[T1]{fontenc}
\usepackage[latin2]{inputenc}
\ifx\pdfoutput 1\usepackage{ae,aecompl}\fi
\def\ifpdf#1#2{\ifx\pdfoutput1{#1}\else{#2}\fi}
\ifpdf{\usepackage{ae}\pdfcompresslevel=9}{}
\begin{document}
\title{Quantum trajectory approach to stochastically-induced quantum interference effects in coherently-driven
  two-level atoms}

\author{A. Karpati}
\author{P. Adam}
\author{W. Gawlik$^{\dagger}$}
\author{B. \L obodzi\'nski$^{\dagger}$\footnote{Present address: DESY Zeuthen,
Platanenallee 6, D-15738 Zeuthen, Germany.}}
\author{J. Janszky}
\affiliation{
Department of Nonlinear and Quantum Optics, \\
  Research Institute for Solid State Physics and Optics, \\ 
  Hungarian Academy of Sciences, P.O. Box 49, H-1525 Budapest, Hungary\\
\\
Institute of Physics, University of P\'ecs, Ifj\'us\'ag \'ut 6.
  H-7624 P\'ecs, Hungary\\
\\
$^\dagger$ Marian Smoluchowski Physical Institute, \\
Jagellonian University, Reymonta 4, 30-059 Krak\'ow, Poland\\
} 
\date{\today}

\begin{abstract}
  Stochastic perturbation of two-level atoms strongly driven by a coherent
  light field is analyzed by the quantum trajectory method.  A new method is
  developed for calculating the resonance fluorescence spectra from numerical
  simulations.  It is shown that in the case of dominant incoherent
  perturbation, the stochastic noise can unexpectedly create phase correlation
  between the neighboring atomic dressed states. This phase correlation is
  responsible for quantum interference between the related transitions
  resulting in anomalous modifications of the resonance fluorescence spectra.
\end{abstract}

\maketitle

\section{Introduction}
Quantum interference is one of the most intriguing phenomena of quantum
mechanics. Over the past decade several effects in atom-light interaction
which have their origin in quantum interference have been predicted and
demonstrated experimentally \cite{arimondo1996}. Some characteristic examples
are reduction and cancellation of absorption
\cite{harris,scully,imamoglu,zhou1, toor} and spontaneous emission
\cite{zhu3,zhu1,xia,zhu2}, and narrow resonances in fluorescence
\cite{zhouswain,zhouswain2}. A prerequisite of quantum interference between
the transition channels is the existence of some stable time correlation of
the atomic system under consideration. A possible way of achieving such
correlation is the application of coherent coupling in a multi-level atomic
system. Although some interference effects have also been found in two-level
systems interacting with two light beams \cite{grynberg}, so far quantum
interference has been observed exclusively in at least three-level systems.

Generally, various incoherent perturbations destroy the phase correlation
between the states involved in the interfering transition pathways and the
coherently induced quantum interference disappears.
However, under special circumstances, even incoherent perturbation can be
responsible for quantum interference. For example,
quantum interference can appear in three-level systems due to collisions. Such
effects are known as pressure-induced extra resonances and have been studied
in four-wave mixing\cite{pr1,pr}. 

Recently, in an experiment with coherently driven two-level atoms, anomalous
resonance fluorescence spectra were found when the collisional relaxation rate
exceeded the Rabi frequency \cite{gawlik}.  The spectra were of the form of a
pressure-broadened line with a narrow, not collisionally broadened dip. These
results, unexpected in a collisionally perturbed two-level system, were
interpreted as a consequence of quantum interference between different
dressed-state transition channels.  In Ref.\cite{gawlik}, it was pointed out
that these effects can also occur in the case of a non-monochromatic, e.g.
phase-diffusing, laser field.  Indeed, resonance fluorescence spectra with the
phase-diffusing laser field have been calculated by Peng Zhou et al.
\cite{zhou2} who obtained the same results as Gawlik et al. \cite{gawlik}.
What in both cases appears essential for observation of quantum interference
and anomalous spectra is that the incoherent perturbation (collisions or phase
diffusion of the light field) dominates over the Rabi oscillations.

These two examples raise an intriguing question how can a stochastic noise 
lead to stable time correlation resulting in quantum
interference in two-level systems.

Quantum trajectory methods are widely used powerful tools for treating the
stochastic evolution of open quantum systems
\cite{carmichael-book,knight-review, garraway2, dalibard1,gardiner1,dum1}. They can provide the solution of
any master equation that is of Lindblad form \cite{lindblad, chien, tian, molmer}. Moreover,
individual quantum 
trajectories, as state evolutions conditioned on particular sequence of
observed events, make it possible to reveal phase correlations in the given
system.

While resonance fluorescence spectra can be adequately modeled by the master
equation, we find the use of quantum trajectories provides more physical
insight. Besides, this method allows to study the stochastic evolution of the
atomic system and, eventually, reveal the phase correlation of its dressed
states.  For that reason, in this paper we analyze in detail the system of
stochastically perturbed two-level atoms applying both the master equation and
the quantum trajectory method to explain quantum interference effects and the
underlying physical processes.

The paper is organized as follows: In Section \ref{sec:model} we introduce our
model for the system of a coherently driven and incoherently perturbed
two-level atom. In Section \ref{sec:numsim} the method of quantum trajectories
is applied to the system and a new method is developed for calculating the
spectrum from the simulation results. Section \ref{sec:fluor} presents the
numerical results for the spectra and compares them with the analytical
solution of the master equation.  In Section \ref{sec:phase} the phase
difference between the dressed states of an atom is analyzed and the phase
correlation is revealed. It is shown that the phase correlation emerges as the
noise magnitude increases and the related quantum interference effect is
discussed.

\section{The model}
\label{sec:model}
The system of incoherently perturbed and coherently driven two-level atoms can
be modeled in several ways. Here we make a rather general assumption that the
stochastic perturbation is responsible for fluctuations of the atomic
resonance frequency which obeys the Gaussian statistics. In particular, such
fluctuations may result from e.g. elastic, dephasing collisions.

In our model the Hamiltonian of the strongly, coherently driven atom subjected
to 
stochastic perturbation has the form
\begin{equation}
\label{eq:hami}
H_{AL}=\hbar(\omega_a(t)-\omega_L) S^z + {1\over 2}\hbar \Omega(S^-+S^+)
\end{equation}
in the interaction picture, where $\omega_a(t) = \omega_{a}+\delta\omega_a(t)$ is the fluctuating atomic
transition frequency, $\omega_L$ is the frequency of the laser, $\Omega$ the
Rabi frequency, and $S^z$, $S^+$, $S^-$ are
the atomic operators defined in the excited state ($|e\rangle$) -- ground
state ($|g\rangle$) basis:
\begin{equation}
S^z={1\over2}\left(\begin{array}{cc}\phantom{-}1&0\\0&-1\end{array}\right),\ 
S^+=\left(\begin{array}{cc}0&1\\0&0\end{array}\right),\ 
S^-=\left(\begin{array}{cc}0&0\\1&0\end{array}\right).
\end{equation}
We assume that the noise in the transition frequency satisfies
\begin{equation}
\label{noise}
\langle\delta\omega_a(t)\delta\omega_a(t')\rangle=2\Gamma \delta(t-t'),
\end{equation}
where $\Gamma$ stands for the magnitude of the stochastic noise. If the
noise is due to collisions, this quantity is the collision rate between the
atoms.
This model can also describe the system of two level atoms driven by a laser
field with fluctuating phase, if the phase drift is
neglected \cite{zhou2}. In such case, $\Gamma$ represents the laser linewidth.

The time-evolution of the system defined by the Hamiltonian in
Eq. (\ref{eq:hami}) is 
described by the following master equation, taking into account also the
spontaneous emission processes:
\begin{equation}
\label{master}
\dot\rho={1\over i\hbar}[\langle H_{AL}\rangle,\rho]+L\rho,
\end{equation}
where
\begin{subequations}
\begin{eqnarray}
L\rho&=&(L\rho)_{\rm sp}+(L\rho)_{\rm st},\\
(L\rho)_{\rm sp}&=&\gamma(-{1\over 2} (S^+S^-\rho + \rho S^+
S^-)+S^-\rho S^+),\\
(L\rho)_{\rm st}&=&4\Gamma(-{1\over 2}(S^zS^z\rho+\rho S^zS^z)+S^z\rho S^z),
\label{eq:Lindbladstoch}
\end{eqnarray}
\end{subequations}
and $\langle H_{AL}\rangle$ is the mean atomic Hamiltonian obtained by averaging over the
stochastic noise of Eq. (\ref{noise}), and $\gamma$ is the natural linewidth of the atom.

It is natural to introduce the dressed-state basis in which the atomic
Hamiltonian $H_{AL}$ is diagonal:
\begin{subequations}
\begin{eqnarray}
\label{dresseda}
|1\rangle&=&\phantom{-}\cos\Theta |g\rangle + \sin\Theta |e\rangle,\\
\label{dressedb}
|2\rangle&=&-\sin\Theta |g\rangle + \cos\Theta |e\rangle,
\end{eqnarray}
where
$$
\Theta = -{1\over 2}\arctan\left({\Omega\over\Delta}\right),
$$
and $\Delta = \omega_a - \omega_L$ is the laser
detuning. 
\end{subequations}

The mean atomic 
Hamiltonian $\langle H_{AL}\rangle$ in the dressed-state basis $|1\rangle$, $|2\rangle$ can
be written as
\begin{equation}
\langle H_{AL}\rangle =E_1 |1\rangle\langle 1| + E_2|2\rangle\langle 2|,
\label{eq:meanHam}
\end{equation}
where
$$
E_{1,2} = \mp {1\over2}\hbar \sqrt{\Omega^2+\Delta^2}.
$$

The effect of incoherent perturbation on the pure states can be determined
from the 
Lindblad form (\ref{eq:Lindbladstoch}) of the master equation in Eq. (\ref{master}).
The action of the operator $2\sqrt{\Gamma}S^z$ corresponds to an event
generated by the stochastic noise.  Without
detuning, this operator generates transitions between the dressed states
$|1\rangle$ and $|2\rangle$: 
\begin{subequations}
\label{eq:dressedtime}
\begin{eqnarray}
2\sqrt{\Gamma}S^z|1\rangle&=& \sqrt{\Gamma}|2\rangle,\\
2\sqrt{\Gamma}S^z|2\rangle&=& \sqrt{\Gamma}|1\rangle.
\end{eqnarray}
\end{subequations}

Another way of observing the effect of the stochastic noise is to transform the
time-dependent Hamiltonian in the Langevin equation 
 into the
dressed-state basis. This must be done carefully since parameter $\Theta$
in the definition of the dressed states becomes time-dependent
in this case:
\begin{subequations}
\begin{eqnarray}
|1,t\rangle \approx |1\rangle - {1\over 2}{\Omega\over
 \Delta^2+\Omega^2}\delta\omega_a(t) |2\rangle,\\
|2,t\rangle \approx |2\rangle + {1\over 2}{\Omega\over
 \Delta^2+\Omega^2}\delta\omega_a(t) |1\rangle
\end{eqnarray}
\end{subequations}
to the first order in $\delta\omega_a(t)$. Hamiltonian $H_{AL}$ is
diagonal in the time-dependent dressed-state basis, thus 
\begin{eqnarray}
\label{eq:dressedham}
H_{AL}&=& E_1 |1,t\rangle \langle 1,t| + E_2 |2,t\rangle \langle 2,t|
\approx\\
&&\approx E_1 |1\rangle \langle 1| + E_2 |2\rangle\langle 2| + \nonumber\\
&&+\delta\omega_a(t)
{1\over 2}{E_1 \Omega\over
 \Delta^2+\Omega^2}(-|1\rangle\langle 2| - |2\rangle \langle 1|)+\nonumber\\
&&+ \delta\omega_a(t)
{1\over 2}{E_2 \Omega\over
 \Delta^2+\Omega^2}(|1\rangle\langle 2| + |2\rangle \langle 1|)
=\nonumber\\
&=& \langle H_{AL}\rangle - \hbar\delta\omega_a(t){\Omega\over
  \sqrt{\Omega^2+\Delta^2}}(|1\rangle\langle 2|+|2\rangle\langle 1|).\nonumber
\end{eqnarray}
This also shows  that the stochastic noise generates transitions between the
dressed states $|1\rangle$ and $|2\rangle$.

 The master equation has the following form in the dressed-state basis:
\begin{subequations}
\begin{eqnarray}
\label{master1}
{\textrm{d}\over \textrm{dt}}{\rho_z}&=&-\left(2{\Gamma'\Omega^2\over
  \Omega^2+\Delta^2}+\gamma\right)\rho_z+\nonumber\\*
&&+2{\Gamma'\Delta\Omega\over
  \Omega^2+\Delta^2}(\rho_{21}+\rho_{12})+\gamma{\Delta\over\sqrt{\Omega^2+\Delta^2}},\\
\label{master2}
{\textrm{d}\over \textrm{dt}}\rho_{12}&=&{\Gamma'\Omega\Delta\over\Omega^2+\Delta^2}\rho_z+{{1\over
  2}\gamma\Omega\over\sqrt{\Omega^2+\Delta^2}}-\nonumber\\*
&&-\left(2\Gamma'{\Delta^2\over\Omega^2+\Delta^2}+i\sqrt{\Omega^2+\Delta^2}+\gamma\right)\rho_{12}-\nonumber\\*
&&-\Gamma'{\Omega^2\over\Omega^2+\Delta^2}(\rho_{12}-\rho_{21}),
\end{eqnarray}
\end{subequations}
where $\rho_z = \rho_{11}-\rho_{22}$, $\Gamma'=\Gamma-\gamma/4$, and
$\rho_{11}$, $\rho_{12}$, $\rho_{21}$, $\rho_{22}$ are the matrix
elements of the density operator in the dressed-state basis. The
matrix element $\rho_{21}$ is the complex conjugate of $\rho_{12}$, as
$\rho$ is Hermitian. For resonant excitation ($\Delta=0$) the
stochastic noise couples the $|1\rangle$ and $|2\rangle$ states and
increases the relaxation rate of the spin $z$ component. In this case
the dressed states become independent of the Rabi frequency, and
Eq.~(\ref{master1}) is uncoupled from Eq. (\ref{master2}). This is
however not the case in the general, non-resonant case.

\section{Numerical simulation}
\label{sec:numsim}
In the system of coherently driven stochastically perturbed two-level atoms,
quantum interference effects can be seen in the resonance fluorescence spectra
\cite{gawlik,zhou2}. The resonance fluorescence can be described by
 transitions between appropriate dressed states of the atom. If spectral
 modifications are due to quantum interference, some time correlation
 should exist between the dressed states of the atom involved in the
 interfering transition channels.

 For analyzing time correlations in a 
 quantum system, quantum trajectory methods are particularly appropriate. 
 These methods are based on the simulation of quantum trajectories, that are
 individual realizations of the evolution of the system conditioned on
 particular sequence of observed events. By tracking the time evolution of a
 single quantum trajectory, the time correlations can be revealed.
 
 We apply the quantum trajectory method of Ref. \cite{dalibard1} for
 simulating the time evolution of the coherently driven, stochastically
 perturbed two-level atom.  In this system, a single quantum trajectory
 evolves coherently according to the Hamiltonian of Eq. (\ref{eq:meanHam}),
 interrupted by incoherent \emph{gedanken} measurements due to noise events
 and spontaneous emission.  The evolution of the density operator of the
 system is obtained by averaging the density operators of the individual
 quantum trajectories. The resulting density operator is the solution of the
 master equation of Eq. (\ref{master}).

The accuracy of the simulation is limited by two factors: the length $\Delta
t$ of the time step and  the number $N$ of the simulated quantum
trajectories. $\Delta t$ should 
be much less than the characteristic time of any process in the system. $N$
should be large enough to obtain the right ensemble averages for the density
operator at the given stochastic noise magnitude. In our simulations $N$ was
approximately $5\cdot 10^5$. 

Within dipole approximation the resonance fluorescence spectrum $S(\omega)$
can be calculated as the real part of the two-time correlation function
\begin{equation}
\Gamma^N_1(\omega)=\lim_{t\rightarrow\infty}\int_0^\infty \exp(-i\omega\tau)\langle S^+(t+\tau)S^-(t)\rangle\, d\tau,
\label{correldef}
\end{equation}
for an arbitrary initial condition:
\begin{equation}
S(\omega)={\rm Re}\Gamma^N_1(\omega),
\end{equation}
where $\omega$ is the detuning of the emitted light from $\omega_L$.
There are different methods in the literature for obtaining the
spectrum using a numerical simulation \cite{dalibard1, tian, molmer}. One kind of them simulates not
only the atom but also the quantized electromagnetic field \cite{tian}. Such methods seem
to be excessive when the field can be treated classically.

The method presented
by Dalibard et al. \cite{dalibard1} simulates only the atom, and obtains the
spectrum by 
calculating two-time averages and taking their Fourier-transform.
 The computation time of this method increases as $1/\Delta t^2$, where
$\Delta t$ is the time step of the simulation, because for each time step an
additional simulation is started to calculate the two-time averages. That can
be time consuming in the case when large number of quantum trajectories 
are simulated and
small time steps are used. This is the situation in our problem when we
simulate the system in the high noise magnitude regime.  

The question arises whether  it is possible to develop a method which
simulates only 
the atom, without the need of starting extra simulations for calculating
the two-time averages. Below, we briefly outline the essentials of our novel
 method for spectrum calculation. More details will be published elsewhere.

Let us consider a general two level atom-field system.
Let $\rho(t)$ be the density operator of the whole system, $A$ an operator  in
the Schr\"odinger picture acting only on the atom, and $U(t)$ the unitary
time-evolution operator. Then
\begin{eqnarray}
{\rm Tr}(A\rho(t))&=&{\rm Tr}(A U(t)\rho(0) U^\dagger(t)) =\nonumber\\
&=& {\rm Tr}(U^\dagger(t)A
U(t)\rho(0)) = {\rm Tr}(A(t)\rho(0)),\nonumber
\end{eqnarray}
where $A(t)$ is the operator $A$ in the Heisenberg picture. One can define
a time-dependent $A'(t)$ operator for which
\begin{equation}
{\rm Tr}(A\rho(t))={\rm Tr}_A(A\rho_A(t)) = {\rm Tr}_A(A'(t)\rho_A(0)),
\label{Avdef}
\end{equation}
where $\rho_A(t) = {\rm Tr}_L \rho(t)$ is the reduced density operator of the
atom. $A'(t)$ depends also on the laser field. 
Let $R_i(0)$ be a set of reduced density operators of the atom that form
a ${\mathbb C}$-linear basis in the set of the operators acting on the
atom.  In the case of a two-level atom the basis consists of four elements.
These basis elements may evolve also, their value at time $t$ is denoted by
$R_i(t)$.
Using this basis, any operator $X$ which acts on the atom can be written in the
form
\begin{equation}
X=\sum_i x_i R_i(0).
\label{Xkifejtes}
\end{equation}
The coefficients can be expressed as
\begin{equation}
x_i = \sum_j (T^{-1})_{ij}{\rm Tr}(X R_j(0)),
\label{lambdaeq}
\end{equation}
where $T_{ij} = {\rm Tr}(R_i(0)R_j(0))$. Matrix $T$ is invertable since
the operators $R_i(0)$ form a basis and they are linearly independent.  Using
Eqs. (\ref{Avdef}), (\ref{Xkifejtes}) and (\ref{lambdaeq})  the following form
can be derived for the $A'(t)$ operator:
\begin{equation}
\label{Avresult}
A'(t) = \sum_{ik}(T^{-1})_{jk}{\rm Tr}(R_k(t)R_i(0))\lambda_i(0) R_j(0)
\end{equation}
where $\lambda_i(0)$ is expressed by the atomic operator $A$ using Eq.~(\ref{lambdaeq}) as
$$
\lambda_i(0) = (T^{-1})_{ij}{\rm Tr}(A R_j(0)).
$$ 
Equation (\ref{Avdef}) holds for all density operators
$\rho(0)$ with $A'(t)$ of Eq. (\ref{Avresult}):
\begin{equation}
{\rm Tr}(A(t)\rho(0)) = {\rm Tr}_A(A'(t)\rho_A(0)).
\label{Avdefheis}
\end{equation}
 Having an operator $B(t)$ in the Heisenberg picture such that
$B(0)$ acts only on the atom, $B(0)\rho(0)$ can 
be expressed as a ${\mathbb C}$-linear combination of density operators. 
The linearity of the trace in Eq. (\ref{Avdefheis}) yields
\begin{equation}
{\rm Tr}\left( A(t)B(0)\rho(0)\right) = {\rm Tr}\left( A'(t) B(0) \rho_A(0)\right).
\end{equation}
For calculating two-time correlation functions of the form 
${\rm Tr}\,A(t)B(t')\rho(0)$, the above equation can be modified by using the cyclic property of the trace:
\begin{eqnarray}
&&{\rm Tr}\,A(t)B(t')\rho(0) = {\rm Tr}\,A(t)U(-t')B(0)U(t')U(-t')\cdot\nonumber\\
&&\quad\cdot \rho(t')U(t') = {\rm Tr}\,U(t')A(t)U(-t')B(0)\rho(t') =\nonumber\\
&&= {\rm Tr}\,A(t-t') B(0)\rho(t') =\nonumber\\
\label{eq:twotime}
&&= {\rm Tr}_A\, A'(t-t') B(0) \rho_A(t'),
\end{eqnarray}
where $\rho_A(t')$ is the reduced density operator in the Schr\"odinger picture
at time $t'$.

Let us apply the general expressions presented above to the atomic operators $S^+$ and
$S^-$. Calculating the correlation function, the quantity ${\rm Tr}\,
S^+(t+\tau)S^-(t)\rho(0)$ should be determined from the simulation. Using Eq.~(\ref{eq:twotime}),
\begin{eqnarray}
&&\lim_{t\rightarrow\infty}{\rm Tr}\, S^+(t+\tau)S^-(t)\rho(0) = \lim_{t\rightarrow\infty}{\rm Tr}\,
S^+(\tau)\cdot\nonumber\\
&&\quad\cdot S^-(0)\rho(t) = {\rm Tr}_A\, {S^+}'(\tau) S^-(0)\rho_A(\infty).
\end{eqnarray}

In order to obtain operator ${S^+}'(\tau)$ from the simulation one needs to
choose a basis consisting of density operators, according to
Eq. (\ref{Xkifejtes}) and start independent
simulations using the elements of this basis as initial states. 
In our simulation we choose the density operators that in the dressed-state
basis $|1\rangle$, $|2\rangle$ are:
\begin{eqnarray}
R_1(0)&=&\left[\begin{array}{rr}1&0\\0&0\end{array}\right],\qquad R_3(0)={1\over 2}\left[\begin{array}{rr}1&-i\\i&1\end{array}\right],\nonumber\\*
R_2(0)&=&{1\over 2}\left[\begin{array}{rr}1&1\\1&1\end{array}\right],\quad
R_4(0)={1\over2}\left[\begin{array}{rr}1&-1\\-1&1\end{array}\right].
\end{eqnarray}
For all time steps of the simulation we calculate the ${S^+}'(\tau)$ operator
using Eq. (\ref{Avresult}) and record it for later use. After the simulation
has been completed, i.e. the time $t$ has reached its final value $T$, the correlation function
defined by Eq. (\ref{correldef}) is calculated numerically by evaluating the
expression
\begin{equation}
\Gamma^N_1(\omega)=\sum_{\tau=0}^{T} \exp(-i\omega\tau){\rm Tr}\, {S^+}'(\tau)S^-(0)
\overline{\rho(T)} \Delta t,
\end{equation}
where the summation is done over all time steps between $0$ and $T$ and
$\overline{\rho(T)}$ is the average of all four $R_i(T)$ density operators.

The advantage of this method is that there is no need to start a new
simulation in each time step, and it is sufficient to simulate only the atomic
system for obtaining the spectra.

\section{The fluorescence spectrum}
\label{sec:fluor}
In order to check our numerical results, we compare them with the spectra
calculated analytically. After determining the time evolution of the averages
of the Block vector components $\langle S^z(t)\rangle$, $\langle
S^+(t)\rangle$, $\langle S^-(t)\rangle$, the quantum regression theorem is
used for expressing the two-time average $\langle S^+(t+\tau)S^-(t)\rangle$ in
Eq.~(\ref{correldef}) as a function of one-time averages \cite{agarwal}.  The
Bloch equations are the following:
\begin{eqnarray}
\dot{\langle S^z(t)\rangle}&=&-\gamma\langle S^z(t)\rangle+{1\over 2}i\Omega(\langle S^-(t)\rangle-\langle S^+(t)\rangle)-{1\over2}\gamma,\nonumber\\
\dot{\langle S^+(t)\rangle}&=&-i\Omega\langle
S^z(t)\rangle+(\phantom{-}i\Delta-2\Gamma-{1\over 2}\gamma)\langle S^+(t)\rangle,\nonumber\\
\dot{\langle S^-(t)\rangle}&=&\phantom{-}i\Omega\langle
S^z(t)\rangle+(-i\Delta-2\Gamma-{1\over 2}\gamma)\langle S^-(t)\rangle.\nonumber
\end{eqnarray}
After some calculation we obtain
\begin{widetext}
\begin{eqnarray}
\Gamma^N_1(\omega)&=&{i\Omega(i\omega-i\Delta+2\Gamma+{1\over2}\gamma)\left(-{1\over
      2}K_3-K_1K_3\right)
\over
\left(i\omega+\gamma\right)\left((i\omega+2\Gamma+{1\over2}\gamma)^2+\Delta^2\right)+\Omega^2(i\omega+2\Gamma+{1\over2}\gamma)
}+\label{correl}\\
&&{
\left({1\over 2}\Omega^2
  +(i\omega+\gamma)(i\omega-i\Delta+2\Gamma+{1\over2}\gamma)\right)\left({1\over 2}+K_1-K_2K_3\right)
+{1\over 2}\Omega^2\left(-K_3K_3\right)
\over
\left(i\omega+\gamma\right)\left((i\omega+2\Gamma+{1\over2}\gamma)^2+\Delta^2\right)+\Omega^2(i\omega+2\Gamma+{1\over2}\gamma)
},\nonumber
\end{eqnarray}
where
\begin{subequations}
\begin{eqnarray}
K_1&=&{
{1\over 2}\gamma\left((2\Gamma+{1\over2}\gamma)^2+\Delta^2\right)
\over
\gamma\left((2\Gamma+{1\over2}\gamma)^2+\Delta^2\right)+\Omega^2(2\Gamma+{1\over2}\gamma)
},\\
K_2&=&{
{1\over 2}\gamma i\Omega\left(i\Delta-2\Gamma-{1\over2}\gamma\right)
\over
\gamma\left((2\Gamma+{1\over2}\gamma)^2+\Delta^2\right)+\Omega^2(2\Gamma+{1\over2}\gamma)
},\\
K_3&=&{
{1\over 2}\gamma i\Omega\left(i\Delta+2\Gamma+{1\over2}\gamma\right)
\over
\gamma\left((2\Gamma+{1\over2}\gamma)^2+\Delta^2\right)+\Omega^2(2\Gamma+{1\over2}\gamma)
}.
\end{eqnarray}
\end{subequations}
For the special case of no detuning
($\Delta=0$) the correlation
function has the form
\begin{eqnarray}
\label{correlspec}
\Gamma^N_{\rm 1}(\omega)&=&{1\over \left(\omega-2i(\Gamma+{1\over 4}\gamma)\right)\left(i\alpha^2(2\omega^2-4i\omega\Gamma''-\alpha^2)\right)}\Bigg[
(4\Gamma+\gamma)\left((\alpha^2+\gamma^2)\Omega^2-2\Omega^4-\alpha^4\right)-2\Omega^6
\\*
&&+\omega^2
(2\alpha^4-2\Omega^2(\alpha^2+\gamma^2))
+i\omega\left(
(3\gamma+4\Gamma'')\Omega^2\alpha^2-4\Gamma''\alpha^4-2\gamma(2\Omega^2-\gamma^2)\Omega^2
\right)\Bigg],\nonumber
\end{eqnarray}
\end{widetext}
where
\begin{equation}
\alpha^2=\gamma^2+4\Gamma\gamma+\Omega^2,\quad \Gamma'' = \Gamma' +
\gamma, \quad \Gamma' = \Gamma - {1\over 4}\gamma.
\end{equation}

The correlation function of Eq. (\ref{correlspec}) can be split into the sum of three
functions:
\begin{equation}
\Gamma^N_{\rm 1}(\omega)={A_+\over \omega-s_+}+{A_-\over \omega-s_-}+{A_0\over\omega-s_0},
\end{equation}
where
\begin{subequations}
\begin{eqnarray}
\label{eq:smin}
s_\pm&=&i\Gamma''\pm i\sqrt{{\Gamma'}^2-\Omega^2},\\
s_0&=& 2i\Gamma'+i\gamma,\\
A_\pm&=&
i\Omega^2{{1\over
    2}\alpha^2+3\Gamma'\gamma\over
2i\alpha^2(s_+-s_-)(is_-+\gamma)}-
\nonumber\\
&&-i\Omega^2{-4\gamma^2\Gamma''(\Gamma+{1\over 4}\gamma)\pm i\gamma\alpha^2(s_+-s_-)
\over
2i(s_+-s_-)\alpha^4(is_-+\gamma)},\nonumber\\
A_0&=&-i{\Omega^2+(4\Gamma+\gamma)\gamma\over 2\alpha^2}.\nonumber
\end{eqnarray}
\end{subequations}
If $\Gamma'<\Omega$, the spectrum has the form
\begin{eqnarray}
S(\omega)&=&{A_0s_0\over
\omega^2+|s_0|^2}+{{\rm Re} A_+\omega-{\rm Re}(A_+s_+^*)\over (\omega +
  \sqrt{\Omega^2-{\Gamma'}^2}) + {\Gamma''}^2} +\nonumber\\*
&&+ {{\rm Re} A_-\omega-{\rm Re}(A_-s_-^*)\over (\omega -
  \sqrt{\Omega^2-{\Gamma'}^2}) + {\Gamma''}^2},
\end{eqnarray}
showing that the centers of the two latter Lorentzians are 
displaced by $\pm\sqrt{\Omega^2-{\Gamma'}^2}$ relative to the laser frequency. Together with the first
Lorentzian at the laser frequency they form the Mollow triplet \cite{mollow1}.

In the other case when $\Gamma'>\Omega$, all the Lorentzians are centered at
zero frequency, corresponding to $\omega_L$ in the Schr\"odinger picture, but one of them has a negative coefficient, resulting in a dip
in the spectrum. 

For the $\Gamma'>\Omega$ case, one obtains the following expression for the spectrum:
\begin{equation}
\label{eq:spectrum}
S(\omega) = {
A_+s_+
\over
\omega^2+|s_+|^2
}
+
{
A_-s_-
\over
\omega^2+|s_-|^2
}
+
{A_0s_0\over
\omega^2+|s_0|^2
},
\end{equation}
where $A_+s_+$ and $A_0s_0$ are always positive and $A_-s_-$ is negative. 

\begin{figure}[htbp]
\includegraphics[angle=-90,width=\textwidth/2-1cm]{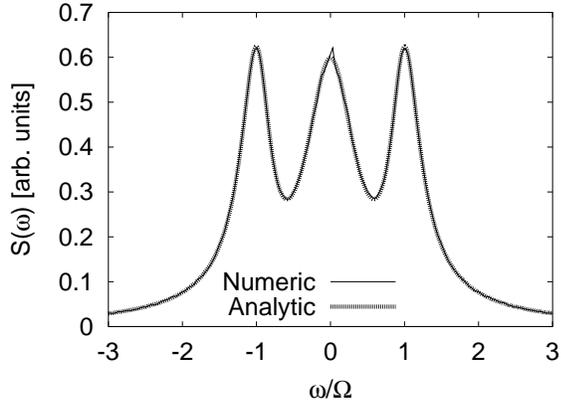}
\caption{The resonance fluorescence spectrum for low noise magnitude and
  strong laser field
  ($\Gamma/\Omega=0.2$, $\gamma/\Omega=0.05$) in the case of no detuning
  ($\Delta=0$).}
\label{fig:3}
\end{figure}

\begin{figure}[htbp]
\includegraphics[angle=-90,width=\textwidth/2-1cm]{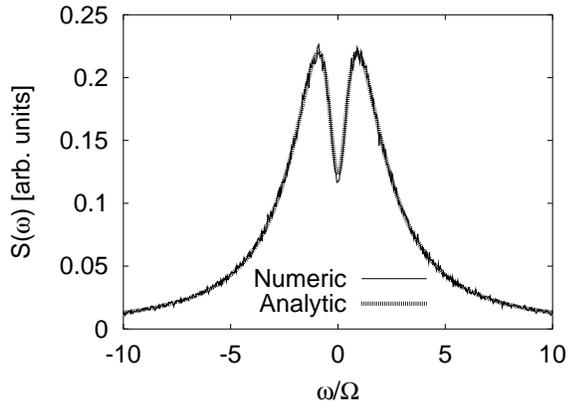}
\caption{The resonance fluorescence spectrum for noise magnitude comparable to
  Rabi frequency and strong laser field ($\Gamma/\Omega=1.1$,
  $\gamma/\Omega=0.05$) in the case of no detuning ($\Delta=0$).}
\label{fig:4}
\end{figure}

\begin{figure}[htbp]
\includegraphics[angle=-90,width=\textwidth/2-1cm]{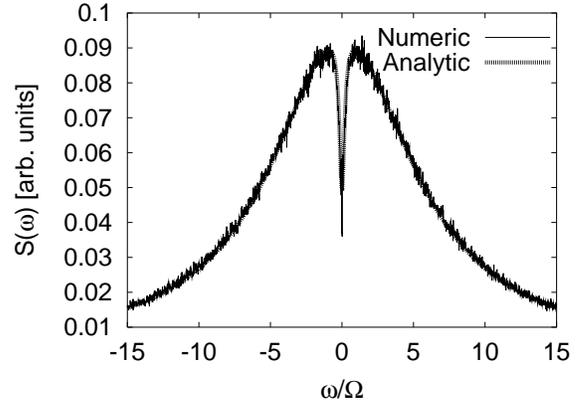}
\caption{The resonance fluorescence spectrum for high noise magnitude and
  strong laser field
  ($\Gamma/\Omega=6$, $\gamma/\Omega=0.05$) in the case of no detuning
  ($\Delta=0$), showing a narrow dip at the center of the spectrum.}
\label{fig:5}
\end{figure}

\begin{figure}[htbp]
\includegraphics[angle=-90,width=\textwidth/2-1cm]{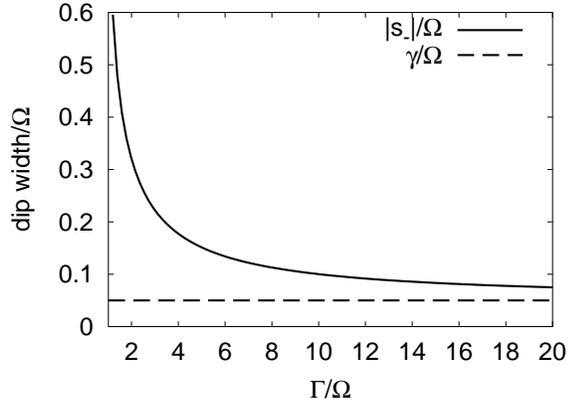}
\caption{The width of the dip in the case of no detuning, i.e. the value of
  $|s_-|$ shown for $\gamma/\Omega=0.05$. When $\Gamma$ increases, the
  width of the dip approaches the natural linewidth $\gamma$.}
\label{fig:sm}
\end{figure}

\begin{figure}[htbp]
\includegraphics[angle=-90,width=\textwidth/2-1cm]{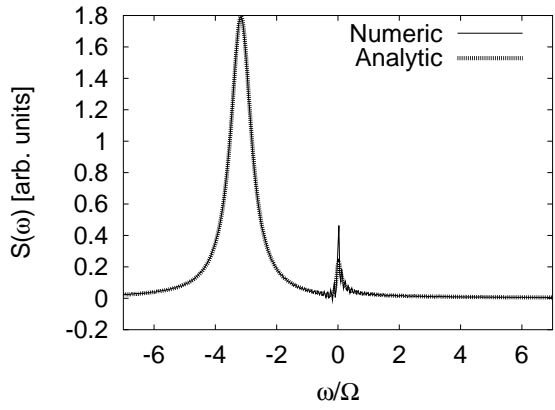}
\caption{The resonance fluorescence spectrum for low noise magnitude,
  strong laser field and large detuning ($\Gamma/\Omega=0.2$,
  $\gamma/\Omega=0.05$, $\Delta/\Omega=3$), showing a Fano-like structure at
  the driving frequency.}
\label{fig:8}
\end{figure}

\begin{figure}[htbp]
\includegraphics[angle=-90,width=\textwidth/2-1cm]{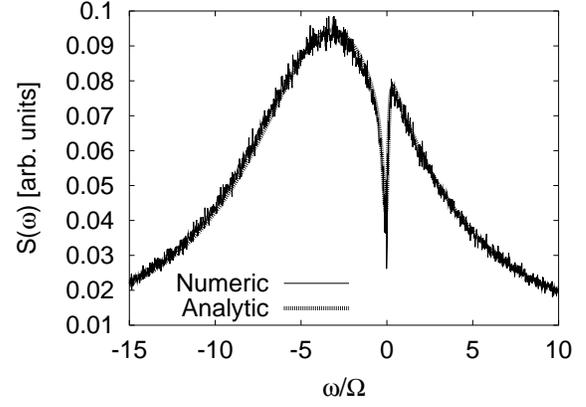}
\caption{The resonance fluorescence spectrum for high noise magnitude,
  strong laser field and large detuning ($\Gamma/\Omega=3$,
  $\gamma/\Omega=0.05$, $\Delta/\Omega=3$), showing an asymmetric Fano profile
  at the driving frequency.}
\label{fig:6}
\end{figure}

In the following we show our numerical results together with the analytical
spectra. Fig.~\ref{fig:3} presents the resonance fluorescence spectrum of the
atom irradiated by a resonant ($\Delta=0$), strong ($\Omega\gg\gamma$) laser
field with low noise ($\Omega>\Gamma$). The spectrum exhibits a three-peak
structure, but with a suppressed and broadened central peak compared to the
standard Mollow triplet. As the noise increases, the central peak disappears
and for $\Gamma$ nearly equal to $\Omega$ we get a two-peak structure with a
relatively broad dip, as depicted in Fig.~\ref{fig:4}.  When the noise
magnitude is much larger than the Rabi frequency, the dip becomes very narrow,
as shown in Fig.~\ref{fig:5}.  The width of the dip is proportional to the
value of the parameter $|s_-|$ of Eq. (\ref{eq:smin}), which approaches the
natural linewidth $\gamma$ when $\Gamma' \gg \Omega$, as shown in Fig.
\ref{fig:sm}.  For large detuning ($\Delta \gg \Omega$) and low noise
($\Gamma\ll\Omega$), a two-peak spectrum is obtained with an asymmetric
Fano-like structure at the center, as depicted in Fig.~\ref{fig:8}.
Increasing the noise magnitude the Fano-like peak transforms to an asymmetric
Fano profile, a narrow dip on the side of the broadened part of the spectrum
due to stochastic noise, and a narrow peak on the other side next to the dip
(Fig.~\ref{fig:6}).

The analytical and numerical results are in very good agreement as one can see
in the presented figures which positively verifies correctness of our
numerical simulations.

\section{Phase correlation and quantum interference}
\label{sec:phase}
The narrow dip in the spectrum (in the case of high noise magnitude, resonant
excitation) and the asymmetric Fano profile (in the case of large detuning and
high noise magnitude) are signatures of quantum interference in the
stochastically perturbed system.  The quantum interference emerges if
long-time phase correlation exists between quantum states connected by
different transition channels. As seen in section \ref{sec:model}, the
stochastic noise generates transitions between the dressed states. It would be
interesting to check whether this coupling has any effect on phase correlation between
them.

\begin{figure}[htbp]
\begin{center}
\boxed{a}\\
\includegraphics[angle=-90,width=\textwidth/2-1cm]{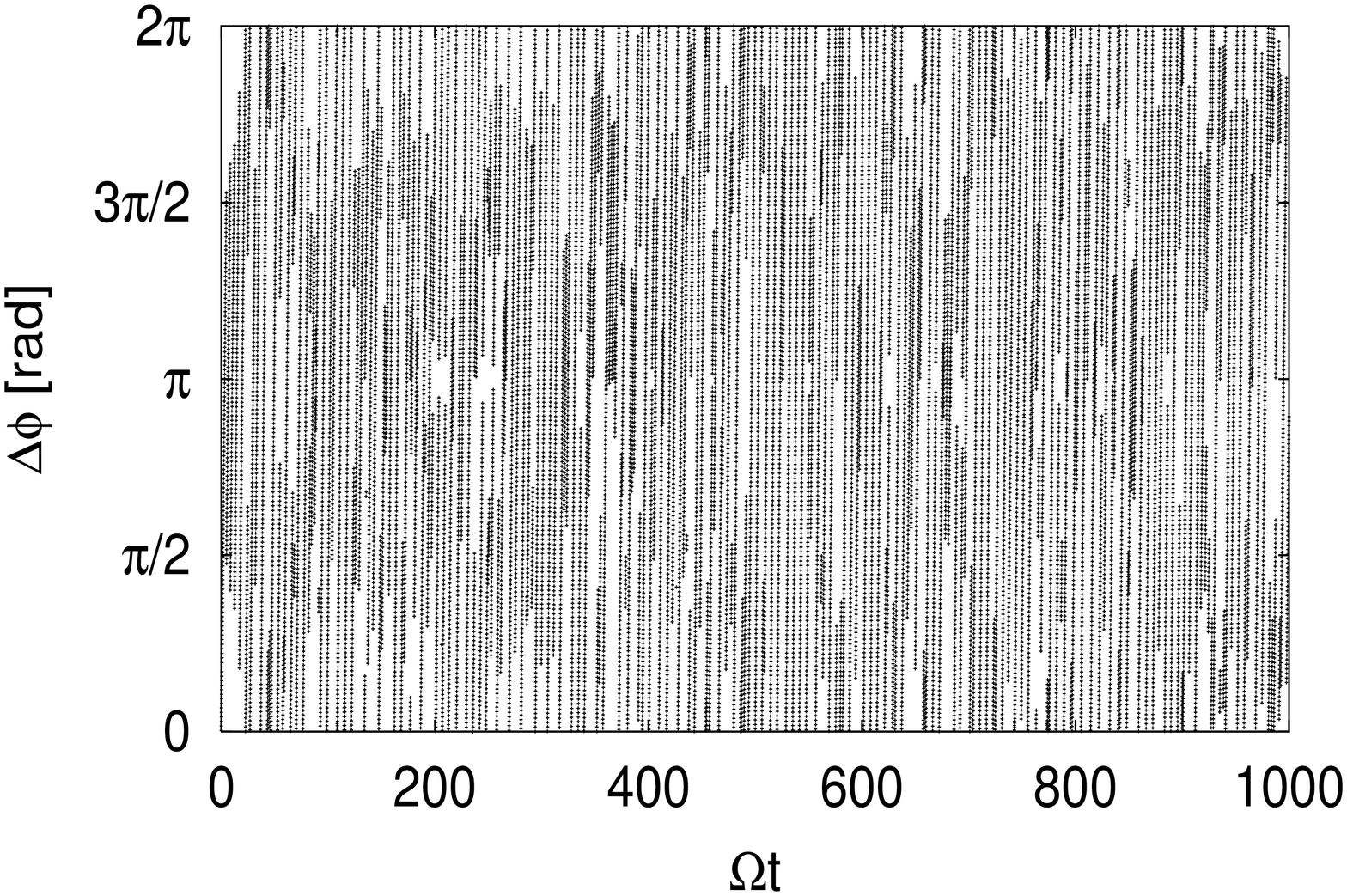}\\[3mm]
\boxed{b}\\
\includegraphics[angle=-90,width=\textwidth/2-1cm]{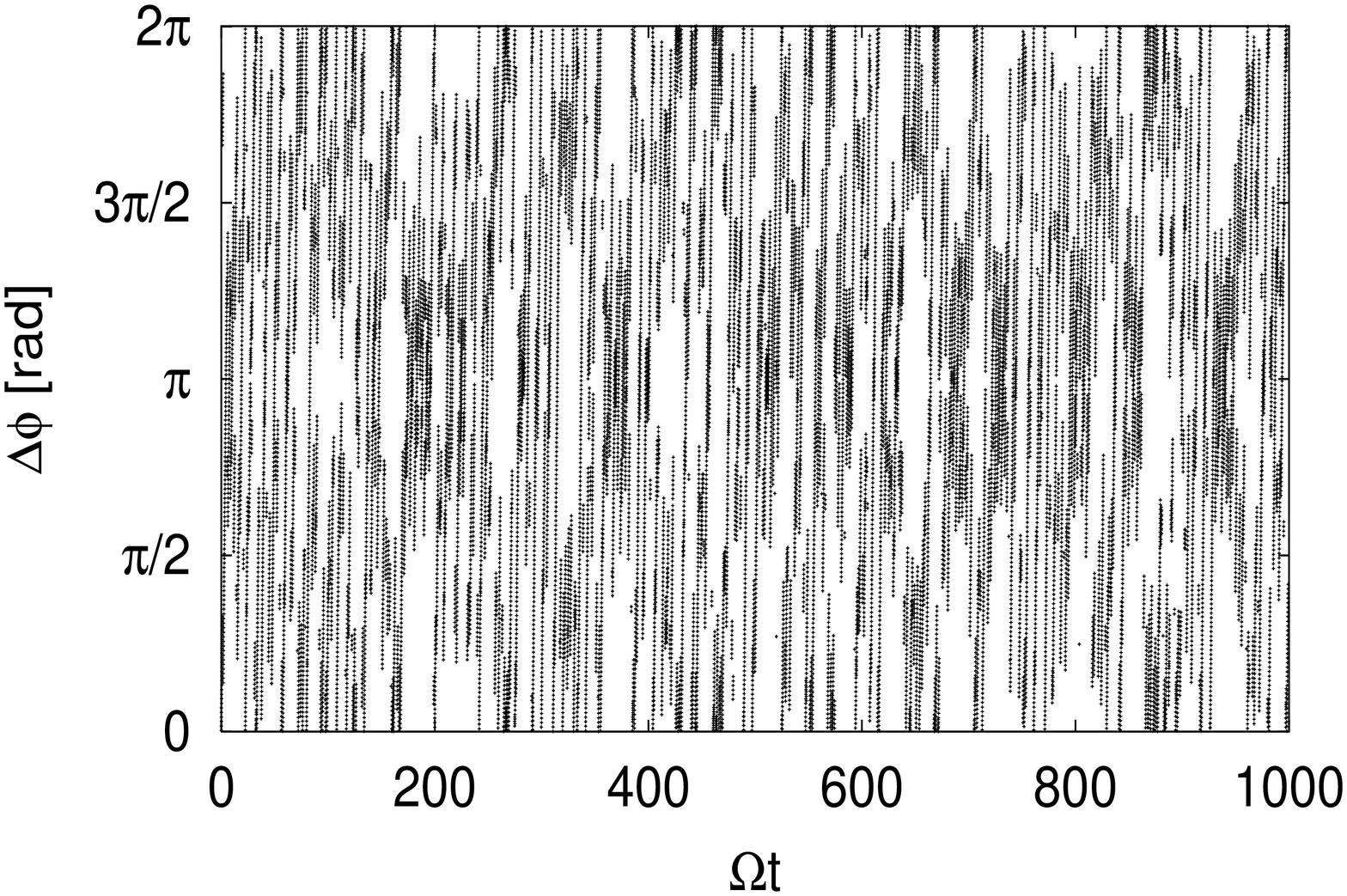}\\[3mm]
\boxed{c}\\
\includegraphics[angle=-90,width=\textwidth/2-1cm]{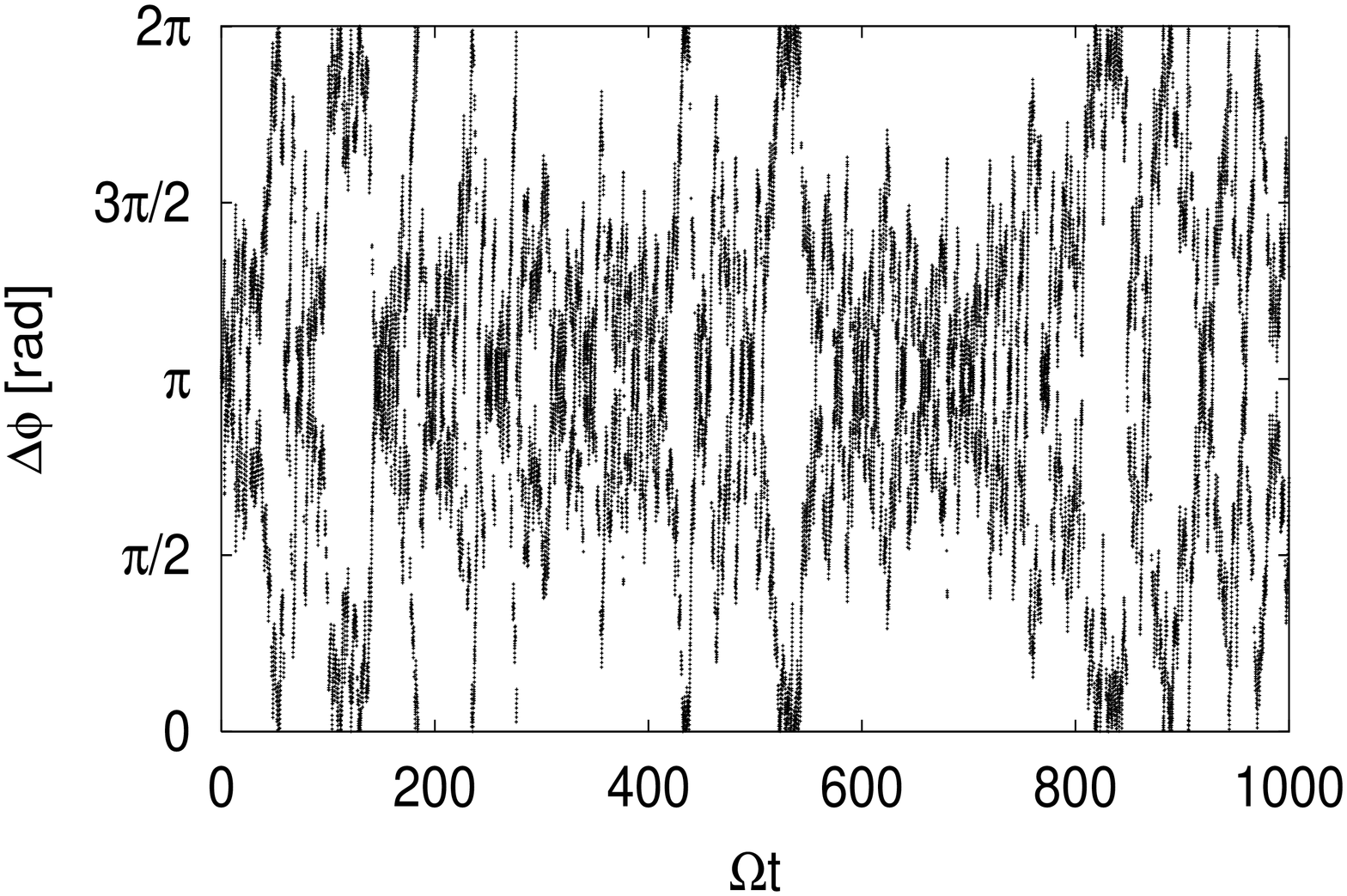}
\end{center}
\caption{
  The phase difference between the dressed states in the case of resonant
  excitation ($\Delta = 0$), a) for a low noise magnitude ($\Gamma/\Omega =
  0.2$, $\gamma/\Omega=0.05$); b) for a noise magnitude comparable to the Rabi
  frequency, ($\Gamma/\Omega = 1.1$, $\gamma/\Omega=0.05$); c) for a high
  noise magnitude, ($\Gamma/\Omega = 5$, $\gamma/\Omega=0.05$). The initial
  state was the excited state $|e\rangle$ in these simulations.}
\label{fig:run1}
\end{figure}

The phase difference $\Delta\phi$ between the dressed states is defined as follows:
\begin{equation}
|\Phi\rangle = a_1e^{i\phi_1}|1\rangle + a_2e^{i\phi_2}|2\rangle,\qquad\Delta\phi = \phi_2-\phi_1,
\end{equation}
where $|\Phi\rangle$ is a pure state of the atom, while $|1\rangle$ and
$|2\rangle$ are the dressed states defined in Eqs.~(\ref{dresseda}) and
(\ref{dressedb}). The phase difference can be calculated straightforwardly
from a single quantum trajectory. It is found that the phase difference
behaves differently in the low and high noise magnitude regimes. In Fig.
\ref{fig:run1}a the noise magnitude is much less than the Rabi frequency, Rabi
oscillations are rarely disrupted by noise events, hence the phase difference
is essentially linearly dependent on time: $\Delta\phi(t)=2\Omega t$.
Consequently, the shape of the phase difference as the function of time shows
no structure. When the noise magnitude increases, as depicted in Fig.
\ref{fig:run1}b, the uniform shape changes to a picture showing some structure
of gaps appearing from time to time between $0$ and $\pi$ values of the phase
difference. For high noise magnitude (Fig. \ref{fig:run1}c), the phase
difference tends to stabilize around values $0$ and $\pi$ for some time
intervals.

In order to  characterize the observed phenomena quantitatively, we
introduce the correlation function of $\cos\Delta\phi$ by the definition
\begin{eqnarray}
C_{\cos}(\tau) &=& c \int_{t=0}^T
(\cos\Delta\phi(t+\tau)-\overline{\cos\Delta\phi})\nonumber\\*
&&\times (\cos\Delta\phi(t)-\overline{\cos\Delta\phi})\, dt,
\end{eqnarray}
where $\overline{\cos\Delta\phi}$ is the mean value of the cosine of
the phase difference for the simulated time interval and $c$ is a
normalization constant fixed by the condition $C_{\cos}(0)=1$.
The correlation function of $\sin\Delta\phi(t)$ is defined similarly.

\begin{figure}[htbp]
\includegraphics[angle=-90,width=\textwidth/2-1cm]{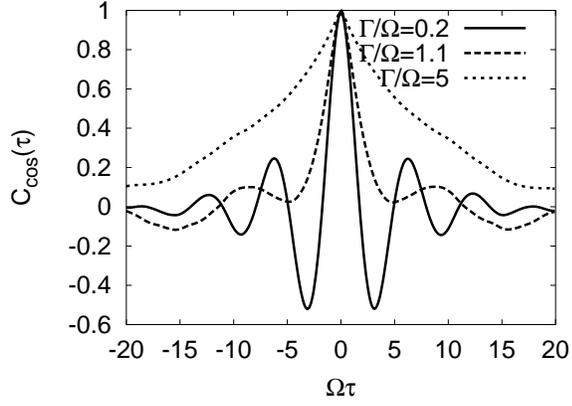}
\caption{Correlation function of $\cos\Delta\phi(t)$  for low,
  medium and high noise magnitude, in the case of no detuning and strong laser
  field ($\Gamma/\Omega\in\{0.2,\,1.1,\,5\}$, $\Delta=0$,
  $\gamma/\Omega=0.05$).}
\label{fig:correl}
\end{figure}

\begin{figure}[htbp]
\includegraphics[angle=-90,width=\textwidth/2-1cm]{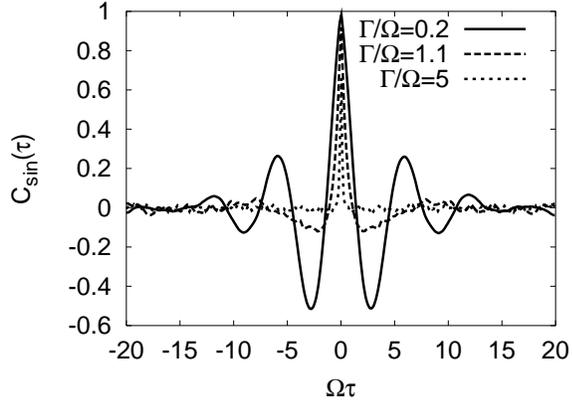}
\caption{Correlation function of $\sin\Delta\phi(t)$  for low,
  medium and high noise magnitude, in the case of no detuning and strong laser
  field ($\Gamma/\Omega\in\{0.2,\,1.1,\,5\}$, $\Delta=0$,
  $\gamma/\Omega=0.05$).  The figure shows that $\sin\Delta\phi(t)$ becomes
  uncorrelated for higher noise magnitude.}
\label{fig:correlsin}
\end{figure}

The $C_{\cos}(\tau)$ function is shown in Fig. \ref{fig:correl} for the same
parameter values as those used in Figs. \ref{fig:run1}a-c. The qualitative
picture of emerging correlations as the noise magnitude increases is now
backed up by the widening of the correlation functions. On the other hand, the
correlation of the sine of the phase difference, $C_{\sin}(\tau)$ (defined
similarly as $C_{\cos}(\tau)$) tends towards a $\delta$-like shape when the
noise increases (Fig.  \ref{fig:correlsin}), so when $\Gamma$ strongly exceeds
$\Omega$, $\sin\Delta\phi(t)$ remains uncorrelated.  This means that the phase
difference is locked to values $0$ and $\pi$ for some time intervals, though
it spans a phase interval no less than ${\pi\over 2}$ around these phase
values.

\begin{figure}[htbp]
\includegraphics[angle=-90,width=\textwidth/2-1cm]{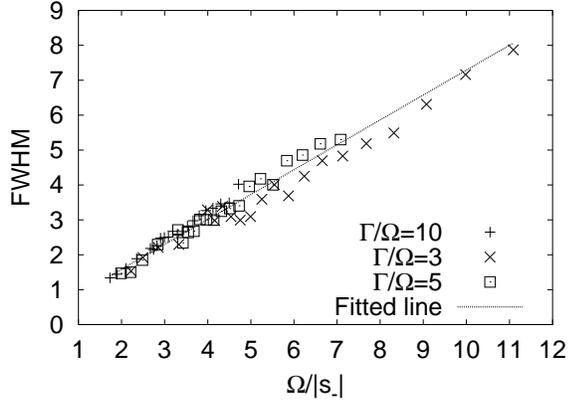}
\caption{The FWHM of the correlation function $C_{\cos}(\tau)$ is
  plotted against the reciprocal of the analytically calculated dip width
  $|s_-|$.}
\label{fig:fwhm}
\end{figure}

It is interesting that the widths of the correlation functions are related
with the observed narrow dips in the spectra. The {\textbf f}ull {\textbf
  w}idth at {\textbf h}alf of the {\textbf m}aximum {\textbf v}alue (FWHM) is
a good measure of the widths of $C_{\rm cos}(\tau)$, and $|s_-|$ as defined in
Eq.~(\ref{eq:smin}) describes well the spectral dip width.
These two quantities are presented in Fig. \ref{fig:fwhm}.
The FWHM of $C_{\cos}(\tau)$ is roughly
proportional to the reciprocal of the width of the dip in the spectrum across
a wide range of parameter sets, so the observed phase correlation
is indeed responsible for the narrowness of the dip in the spectrum.

\begin{figure}[htbp]
\includegraphics[angle=-90,width=\textwidth/2-1cm]{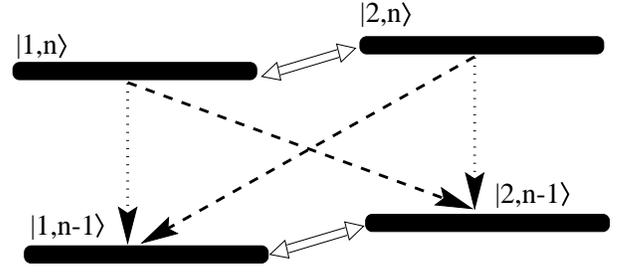}
\caption{Different dressed-state transition channels for a coherently driven
and  stochastically perturbed two-level atom.}
\label{fig:levels}
\end{figure}

The stabilization of the phase difference between the dressed states of the
stochastically perturbed and coherently driven two-level atom is the underlying
physical process which makes the quantum interference possible. This stabilization
supports the following interpretation first suggested for collisional and
phase noise-induced quantum interference effects in resonance fluorescence
spectrum in Ref. \cite{gawlik}.
Resonance fluorescence of a strongly-driven two-level atom is emitted in
cascade transitions downward the ladder of the dressed-state
doublets. Fig. \ref{fig:levels} shows two adjacent doublets and all possible
spontaneous and noise-induced transitions between the dressed-atom states.
According to Eqs. (\ref{eq:dressedtime}) and (\ref{eq:dressedham}), noise
events generate transitions between the dressed states $|1\rangle$ and
$|2\rangle$ and couple them as indicated by double arrows in Fig.
\ref{fig:levels}.  As we have seen in the previous section, in the
noise-dominated regime, i.e. when $\Gamma>\Omega$, the phase difference
between dressed doublets tends to stabilize for some time intervals due to
frequent noise events. Moreover, according to Eq. (\ref{eq:spectrum}), the
resonance frequencies of all fluorescence contributions are the same in this
regime. Among several possible emission channels there are two pairs:
$|1,n\rangle\rightarrow |2,n-1\rangle \rightarrow |1,n-1\rangle$ and
$|1,n\rangle\rightarrow|2,n\rangle \rightarrow |1,n-1\rangle$ (or
$|2,n\rangle\rightarrow |1,n\rangle \rightarrow |2,n-1\rangle$ and
$|2,n\rangle\rightarrow|1,n-1\rangle \rightarrow |2,n-1\rangle$) that differ
exclusively by time ordering between collisional mixing and photon emissions.
Photons emitted along these channels are indistinguishable, so their
interference is possible.  Due to opposite signs of the relevant matrix
elements this interference is destructive and creates a dip in the line
center. On the other hand, other emission channels are not that equivalent,
hence the corresponding photons cannot interfere and contribute to non-zero
intensity at $\omega=0$.  This interference is similar to that seen by Schrama
et al. \cite{schrama} in photon correlations of the well resolved Mollow
triplet components in the opposite limit when $\Gamma < \Omega$.

\section{Conclusion}
We have applied the quantum trajectory method to the system of two-level atoms
strongly driven by a coherent light field and perturbed by stochastic noise.
We have developed a new method for obtaining the resonance fluorescence
spectra from numerical simulations. This method is especially advantageous for
physical systems where the noise dominates the dynamics, and one needs to
simulate many quantum trajectories using small time steps.  The simulation of
a single quantum trajectory revealed that for high noise magnitude the phase
difference between the dressed states tends to stabilize around fixed values.
When calculating the resonance fluorescence spectra, narrow resonances as
central dip and dispersive Fano-like profile occurred in the regime where the
noise dominated the Rabi oscillations.  These modifications of the resonance
fluorescence spectra are associated with the stabilization of the
dressed-state phases and stochastically-induced quantum interference between
various emission channels.

\section*{Acknowledgements}
This work was supported by the Research Fund of Hungary under contract No.
T034484 and by the Polish Committee for Scientific Research (grant 2PO3B 015
16). It was also a part of a general program of the National
Laboratory of AMO Physics in Toru{\'n}, Poland (PBZ/KBN/032/P03/2001).

\clearpage%

\clearpage%

\clearpage

\clearpage%

\clearpage

\end{document}